\documentstyle[epsfig,longtable]{aipproc}

\begin{document}
\title{Super-LOTIS\\ A High-Sensitive Optical Counterpart Search Experiment}

\author{H. S. Park$^1$, E. Ables$^1$, D. L. Band$^5$, S. D. Barthelmy$^3$, 
R. M. Bionta$^1$,  P. S. Butterworth$^3$, T. L. Cline$^3$, D. H. Ferguson$^6$, 
G. J. Fishman$^4$, N. Gehrels$^3$, D. Hartmann$^2$, K. Hurley$^7$, 
C. Kouveliotou$^4$, C. A. Meegan$^4$, L. Ott$^1$, E. Parker$^1$, G. G. Williams$^2$}
\address{$^1$Lawrence Livermore National Laboratory, Livermore, CA 94550\\
$^2$Dept. of Physics and Astronomy, Clemson University, Clemson, SC  29634-1911\\
$^3$NASA/Goddard Space Flight Center, Greenbelt, MD 20771\\
$^4$NASA/Marshall Space Flight Center, Huntsville, AL 35812\\
$^5$CASS 0424, University of California, San Diego, La Jolla, CA 92093\\
$^6$Dept. of Physics, California State University at Hayward, Hayward, CA 94542\\
$^7$Space Sciences Laboratory, University of California, Berkeley, CA 94720-7450}

%\lefthead{LEFT head}
%\righthead{RIGHT head}
\maketitle

\begin{abstract}
We are constructing a 0.6 meter telescope system to search for early time
gamma-ray burst(GRB) optical counterparts. Super-LOTIS (Super-Livermore
Optical Transient Imaging System) is an automated telescope system that has a
0.8 x 0.8$^\circ$ field-of-view, is sensitive to Mv $\sim$ 19 and responds 
to a burst
trigger within 5 min. This telescope will record images of the gamma-ray burst
coordinates that is given by the GCN (GRB Coordinate Network). A measurement
of GRB light curves at early times will greatly enhance our understanding of
GRB physics. 
\end{abstract}

\section*{Introduction}
The origin and nature of gamma-ray bursts (GRBs) remains an important
unresolved problem in astrophysics. GRBs are brief bursts ($<$ 100
sec duration) of high-energy radiation that appear at random in the sky. Much
of the difficulty in studying gamma-ray bursts results from the poor
directional precision (1 $\sim$ 10$^\circ$ 1-$\sigma$ statistical error) 
available from current
gamma-ray detection experiments and their short duration (1 $\sim$ 100 sec). 
Even though recent fading x-ray, optical and radio counterpart 
observations by the
Italian-Dutch satellite (BeppoSAX) \cite{Costa97a}, \cite{Costa97b}, 
\cite{Heise97}
provided information on their distance scale \cite{Metzger97}, these
observations were made many hours later than the GRB. These afterglows may be
due to different process from the GRB production mechanism. An observation of
optical activity simultaneous to the GRB may provide clues to understanding
this process.

In an attempt to search for simultaneous optical counterparts of GRBs, we
initially utilized an existing wide-field-of-view telescope at Lawrence
Livermore National Laboratory (LLNL) to rapidly image GRB coordinates
distributed by the BATSE real-time coordinate distribution network \cite{Bart98}.
This first experiment, the Gamma Ray Optical Counterpart Search Experiment
(GROCSE),
did not find optical optical counterparts at the Mv $\sim$ 7.5 sensitivity level
\cite{Park97a}. 
 
Subsequently, we constructed the Livermore Optical Transient Imaging System
(LOTIS) which has a 17.4 x 17.4$^\circ$ field of view to image the entire error
circle of the rapid GCN notice (BATSE-Original trigger coordinates 
arrive in 5 sec but have $\sim$ 15$^\circ$ errors). 
LOTIS has been operating since Oct. 1996 and we
have not yet observed any simultaneous optical activity at Mv $\sim$ 11 level 
\cite{Park97b}, \cite{Willi98}.
 
In order to provide better GRB coordinates to search for counterparts, the GCN
has installed new triggers called LOCBURST and RXTE which utilize the best
analysis performed by the BATSE team (involving an operator's interaction) and
the RXTE satellite which uses the hard x-ray afterglow to calculate a better
position. The LOCBURST trigger error is 0.2 $\sim$ 2$^\circ$; and the RXTE 
trigger error is 6 $\sim$ 40 arcmin depending on the statistics. 
Although delayed (15 $\sim$ 35 min. for LOCBURST; 3 $\sim$ 5 hr. for RXTE 
triggers), the smaller error box enables a more
conventional, deeper telescope (larger aperture but smaller field of view than
LOTIS) to follow up on the GRBs. Since the bursts are random, this telescope
will need to be dedicated and automated to always be ready for new triggers.
 
In an attempt to make such observations, we are constructing a large aperture
telescope system dedicated to this search.

\section*{SUPER-LOTIS}

\begin{figure}[b!] % fig 1
\centerline{\epsfig{file=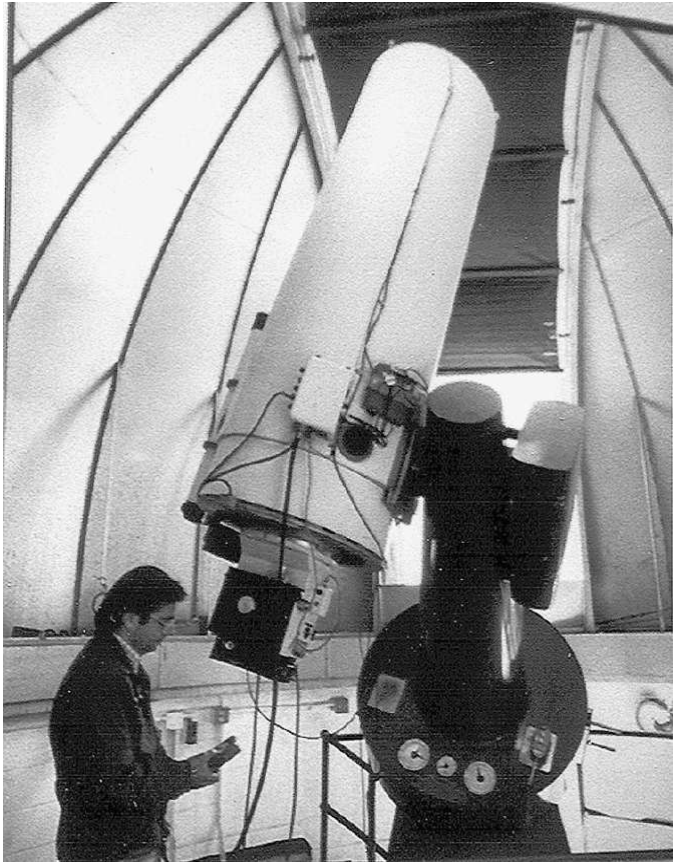,height=4.5in,width=3.5in}}
\vspace{10pt}
\caption{Super-LOTIS Boller and Chivens 0.6 meter reflective telescope.  After
refurbishing by adding computer controlled motors and installing a CCD camera,
this automated system will be dedicated to the GRB optical counterpart search.}
\label{fig1}
\end{figure}

The telescope is a Boller and Chivens 0.6 meter reflective telescope with f/3.5.
Figure \ref{fig1} shows the telescope. It has superb optical quality 
and mechanical
structure; however, it is not equipped with computer controllable drives nor
an electro-optical imaging sensor. We are converting this telescope to
Super-LOTIS by refurbishing the motor drive, installing a CCD camera, and
placing it at a remote site for dedicated observation. As for the sensor, we
are installing a LOTIS CCD camera which utilizes a Loral 442A 2048 x 2048 CCD
(15 x 15 $\mu$m pixels) with LLNL built readout electronics. The CCD is cooled by
thermo electric cooling (to -30$^\circ$C) which minimizes dark current 
and readout noise.
 
The Super-LOTIS will have 0.84 x 0.84$^\circ$ field-of-view 
(1.5 arcsec/pixel) which
is sufficient for BATSE/RXTE trigger types distributed by the GCN. We plan to
dither around the GCN "Original" trigger  coordinates which has only a 5 sec
delay, but a large 15$^\circ$ error box. When we receive refined positions, i.e. 
LOCBURST, or  RXTE triggers, we will scan the region and stay at that location
the rest of the night. Our scanning strategy and automation will allow us to
record GRB optical activity as early as a few minutes.
 
The basic on-line software has already been written and has been operating on
LOTIS. We will need very minimal modification to the existing software for the
entire data acquisition control.

\begin{figure}[b!] % fig 2
\centerline{\epsfig{file=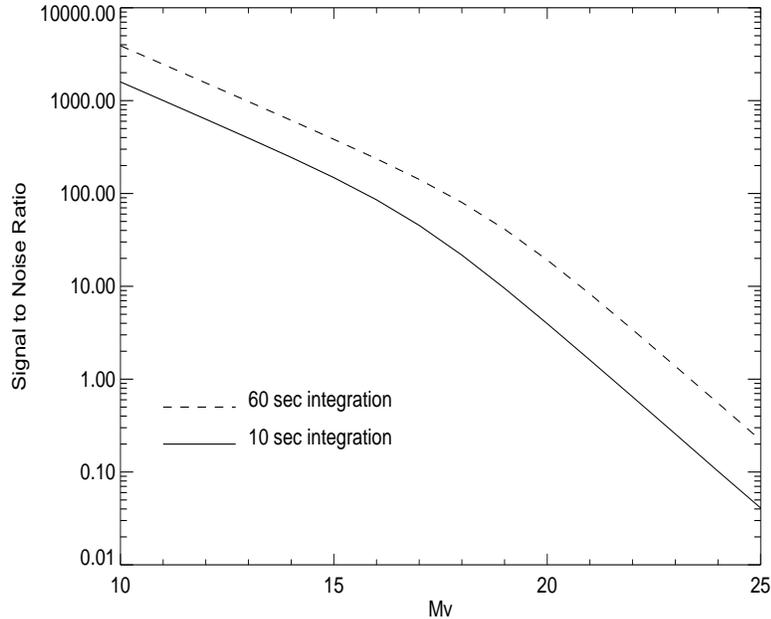,height=3.5in,width=4.0in}}
\vspace{10pt}
\caption{Super-LOTIS sensitivity: Predicted signal to noise ratio vs. visual
magnitude. Super-LOTIS will be able to detect Mv $\sim$ 19 objects with 10 sec
and Mv $\sim$ 21 with 60 sec integration times.}
\label{fig2}
\end{figure}
 
We have estimated the sensitivity of the Super-LOTIS system. The
calculation includes the measured camera dark current at -30$^\circ$C, 
readout noise, typical sky background at an observing site and 
shot noise. Figure \ref{fig2} shows the
resulting signal to noise ratio with 10 and 60 sec integration 
time vs. the visual magnitude. The calculation indicates that the 
Super-LOTIS will see Mv $\sim$ 19 stars at a signal to noise ratio 
of 10 with 10 sec and Mv $\sim$ 21 with 60 sec integration times.

Utilizing our successful experience in construction, operation, data handling,
and data analysis of the GROCSE and LOTIS systems we expect Super-LOTIS to be
constructed and operational within a year. Detection of optical emission at
early times (or placing stringent constraints) would provide a crucial link
between the multiwavelengths properties of the burst and its afterglow. Burst
and afterglow emission are likely to probe different aspects of the
GRB model (e.g. internal vs. external shocks.) So far only two afterglows
have been detected and no optical detection simultaneous with
or shortly after the gamma-ray burst has been made. The 90 minute 
delayed emission of high
energy photons from GRB940217 \cite{Hurley94} suggests that similar emissions in
the optical wavelength could accompany some bursts. Super-LOTIS would
detect such a new spectral component of the bursts to a magnitude level of
$>$ Mv $\sim$ 19. While upper limits will be useful for constraining the models,
Super-LOTIS will establish the GRB light curves at early times which will
provide a crucial step toward understanding GRB phenomenon.


\begin{references}
\bibitem{Costa97a}Costa, E., et al., {\it IAU Circ.}, 6572 (1997).
\bibitem{Costa97b}Costa, E., et al., {\it IAU Circ.}, 6649 (1997).
\bibitem{Heise97}Heise, J., et al., {\it IAU Circ.}, 6654 (1997).
\bibitem{Metzger97}Metzger, M., et al., {\it IAU Circ.}, 6655 (1997).
\bibitem{Bart98}Barthelmy, A., et al., {\it These Proceedings} (1998).
\bibitem{Park97a}Park, H., et al., {\it AstroPhys. Journal}, {\bf 490} (1997).
\bibitem{Park97b}Park, H., et al., {\it AstroPhys. Journal Letters}, {\bf 490}, L21-L24 (1997).
\bibitem{Willi98}Williams, G., {\it These Proceedings} (1998). 
\bibitem{Hurley94}Hurley, K., et al., {\it Nature.}, {\bf 372}, 652 (1994).
\end{references}
\end{document}